\documentclass[aps,prl,superscriptaddress,prbib,11pt]{revtex4}
\usepackage{graphicx}
\usepackage{lmodern}
\usepackage{float}
\usepackage{amssymb}
\usepackage{amsmath}
\usepackage{color}
\usepackage[normalem]{ulem}
\usepackage{cancel}

\newcommand{\SRO}{Sr\textsubscript{2}RuO\textsubscript{4}}

\definecolor{darkred}{rgb}{0.85, 0, 0}
\definecolor{darkgreen}{rgb}{0, 0.65, 0}

\begin{document}

\author{Vadim Grinenko} 
\email{v.grinenko@ifw-dresden.de}
\thanks{These authors contributed equally.}
\affiliation{Institute for Solid State and Materials Physics, Technische Universit\"{a}t Dresden, D-01069 Dresden, Germany}
\affiliation{IFW Dresden, Helmholtzstrasse 20, D-01069 Dresden, Germany}
\author{Shreenanda Ghosh}
\thanks{These authors contributed equally.}
\affiliation{Institute for Solid State and Materials Physics, Technische Universit\"{a}t Dresden, D-01069 Dresden, Germany}
\author{Rajib Sarkar}
\affiliation{Institute for Solid State and Materials Physics, Technische Universit\"{a}t Dresden, D-01069 Dresden, Germany}
\author{Jean-Christophe Orain}
\affiliation{Laboratory for Muon Spin Spectroscopy, Paul Scherrer Institute, CH-5232 Villigen PSI, Switzerland}
\author{Artem Nikitin}
\affiliation{Laboratory for Muon Spin Spectroscopy, Paul Scherrer Institute, CH-5232 Villigen PSI, Switzerland}
\author{Matthias Elender}
\affiliation{Laboratory for Muon Spin Spectroscopy, Paul Scherrer Institute, CH-5232 Villigen PSI, Switzerland}
\author{Debarchan Das}
\affiliation{Laboratory for Muon Spin Spectroscopy, Paul Scherrer Institute, CH-5232 Villigen PSI, Switzerland}
\author{Zurab Guguchia}
\affiliation{Laboratory for Muon Spin Spectroscopy, Paul Scherrer Institute, CH-5232 Villigen PSI, Switzerland}
\author{Felix Br\"uckner}
\affiliation{Institute for Solid State and Materials Physics, Technische Universit\"{a}t Dresden, D-01069 Dresden, Germany}
\author{Mark E. Barber}
\author{Joonbum Park}
\affiliation{Max Planck Institute for Chemical Physics of Solids, D-01187 Dresden, Germany}
\author{Naoki Kikugawa} 
\affiliation{National Institute for Materials Science, Tsukuba 305-0003, Japan}
\author{Dmitry A. Sokolov}
\affiliation{Max Planck Institute for Chemical Physics of Solids, D-01187 Dresden, Germany}
\author{Jake S. Bobowski}
\affiliation{Department of Physics, Kyoto University, Kyoto 606-8502, Japan}
\affiliation{Physics, University of British Columbia, Kelowna, BC V1V 1V7, Canada}
\author{Takuto Miyoshi}
\affiliation{Department of Physics, Kyoto University, Kyoto 606-8502, Japan}
\author{Yoshiteru Maeno}
\affiliation{Department of Physics, Kyoto University, Kyoto 606-8502, Japan}
\author{Andrew P. Mackenzie}
\affiliation{Max Planck Institute for Chemical Physics of Solids, D-01187 Dresden, Germany}
\affiliation{Scottish Universities Physics Alliance (SUPA), School of Physics and Astronomy, University of St. Andrews, St. Andrews KY16 9SS, United Kingdom}
\author{Hubertus Luetkens} 
\affiliation{Laboratory for Muon Spin Spectroscopy, Paul Scherrer Institute, CH-5232 Villigen PSI, Switzerland}
\author{Clifford W. Hicks} 
\email{hicks@cpfs.mpg.de}
\affiliation{Max Planck Institute for Chemical Physics of Solids, D-01187 Dresden, Germany}
\author{Hans-Henning Klauss}
\email{henning.klauss@tu-dresden.de}
\affiliation{Institute for Solid State and Materials Physics, Technische Universit\"{a}t Dresden, D-01069 Dresden, Germany}

\title{Split superconducting and time-reversal symmetry-breaking transitions, and magnetic order in \SRO{} under uniaxial stress}

%\VG{Vadim.} \SG{Shreenanda.} \RS{Rajib.} \CH{Cliff.} \HK{Henning.}

\date{\today}

\begin{abstract}

Among unconventional superconductors, \SRO{} has become a benchmark for experimentation and theoretical analysis because its normal-state electronic structure is known with exceptional precision, and
because of experimental evidence that its superconductivity has, very unusually, a spontaneous angular momentum, \textit{i.e.} a chiral state.  This hypothesis of chirality is however difficult to
reconcile with recent evidence on the spin part of the order parameter.  Measurements under uniaxial stress offer an ideal way to test for chirality, because under uniaxial stress the superconducting
and chiral transitions are predicted to split, allowing the empirical signatures of each to be identified separately. Here, we report zero-field muon spin relaxation (ZF-$\mu$SR) measurements on
crystals placed under uniaxial stresses of up to 1.05~GPa. We report a clear stress-induced splitting between the onset temperatures of superconductivity and time-reversal symmetry breaking,
consistent with qualitative expectations for chiral superconductivity. We also report the appearance of unexpected bulk magnetic order under a uniaxial stress of $\sim$ 1.0~GPa  in clean \SRO{}.

\end{abstract}

\maketitle

\section{Introduction}

Superconductors are classified as conventional or unconventional on the basis of whether the phase of the order
parameter is isotropic in momentum space or not. Unconventional superconductivity, with sign changes in the order
parameter, can have higher critical temperatures because it can be mediated by repulsive interactions that do not work
in opposition to Coulomb repulsion. A very small fraction of unconventional superconductors break time reversal
symmetry, in other words, have imaginary components in their order parameters. Introducing imaginary components can
increase the condensation energy by filling in gap nodes, but also frustrates the pairing interaction: scattering
between sections of Fermi surface where the phase of the order parameter differs by $\pi/2$ does not contribute to the
condensation energy.  Because superconductivity is exponentially sensitive to scattering strength, it is in general
surprising when time-reversal symmetry breaking (TRSB) order parameters are favored. The origin of the superconductivity
in one famous candidate for TRSB superconductivity, \SRO{}, remains a mystery despite 25 years of intense research
\cite{MaenoNature94}.

There is evidence that the superconductivity of \SRO{} is chiral. A nonzero Kerr rotation is observed below the critical
temperature $T_\text{c}$~\cite{Xia06}, and the phenomenology of junctions between \SRO{} and conventional
superconductors offers compelling evidence for domains in the superconducting state~\cite{Kidwingira06, Nakamura12,
Anwar13}. For most of the history of \SRO{} evidence for chirality has been understood in terms of an odd-parity
order parameter with equal spin pairing in the RuO$_2$ planes: $p_x \pm ip_y$~\cite{Mackenzie03, Maeno12, Kallin12,
Mackenzie17}. More recent evidence, especially a revision in NMR data, points to even-parity pairing, as in the vast
majority of known superconductors~\cite{Pustogow19, Ishida19, Steppke17}. Reconciling even parity with chirality
and the tetragonal lattice symmetry of \SRO{} compels consideration of a $d_{xz} \pm id_{yz}$ order parameter.  Under
conventional understanding this is an unexpected order parameter, because the line node at $k_z=0$ appears to imply
pairing of carriers \textit{between} layers, while \SRO{} is a layered metal with very low interlayer
conductivity~\cite{Mackenzie03}. Indeed, the evidence on chirality is mixed. A recent junction experiment finds
time-reversal invariant superconductivity~\cite{Kashiwaya19}. Quasiparticle interference data suggest a $d_{x^2-y^2}$
gap structure~\cite{Sharma19}.  Evidence for chirality has not been found in thermodynamic quantities: as illustrated in
Fig.~1, under a hypothesis of chirality $T_\text{c}$ and $T_\text{TRSB}$, the onset temperature of time-reversal
symmetry breaking, are predicted to split under uniaxial stress~\cite{Sigrist91}, but evidence for this splitting has
not been resolved in either the stress dependence of $T_\text{c}$ or in heat capacity data under uniaxial
stress~\cite{Hicks14Science, Watson18, Li19}.

However there is no widely-accepted alternative hypothesis to understand the experiments that do indicate chirality.
Since a confirmation of chirality may imply a new pairing mechanism it is an important point to resolve. 
The approach taken here is to use a non-thermodynamic probe specifically sensitive to time reversal symmetry
breaking, muon spin relaxation ($\mu$SR), to test for transition splitting in samples placed under uniaxial stress. In
this method, spin-polarized muons are implanted, and each muon spin then precesses in its local magnetic field.
The time evolution of the polarization is determined by collecting statistics on the direction of positron emission as
the muons decay. Like neutron scattering, $\mu$SR is a true bulk-sensitive probe. In contrast to neutron scattering, it
offers sensitivity to spatially uncorrelated fields. The minimum detectable field is 0.01--0.1~mT, against $\sim$1~mT
for neutron scattering. In a small but growing number of known superconductors, muon spin polarization relaxes more
quickly below $T_\text{c}$.  This indicates onset of internal magnetic fields, and is interpreted as a signature of TRSB
superconductivity.  This signal is consistently seen in \SRO{}, where the magnitude of the increase indicates a
superconductivity-related internal field of $\sim$0.05~mT~\cite{Luke98, Luke00, Shiroka12, Higemoto14}.

Our experiment is a probe of the $\mu$SR technique as well as of \SRO{}.  The internal field is thought to arise at
edges, defects, and domain walls: persistent currents are predicted to appear when spatial variation is imposed upon a
TRSB order parameter~\cite{Sigrist91}. This hypothesis is called into question by the fact that edge fields were not
detected, to a sensitivity of $\sim$0.1~$\mu$T, in scanning SQUID microscopy measurements, neither in \SRO{} nor in
another superconductor where $\mu$SR measurements indicate an internal field, PrOs$_4$Sb$_{12}$~\cite{Kirtley07,
Hicks10}. As $\mu$SR is the primary tool by which TRSB superconductivity has been identified in a number of other
materials, clear demonstration that enhanced muon spin relaxation is not an occasional artefact of conventional
superconducting transitions is vital.

\begin{figure}
\includegraphics[width=57mm]{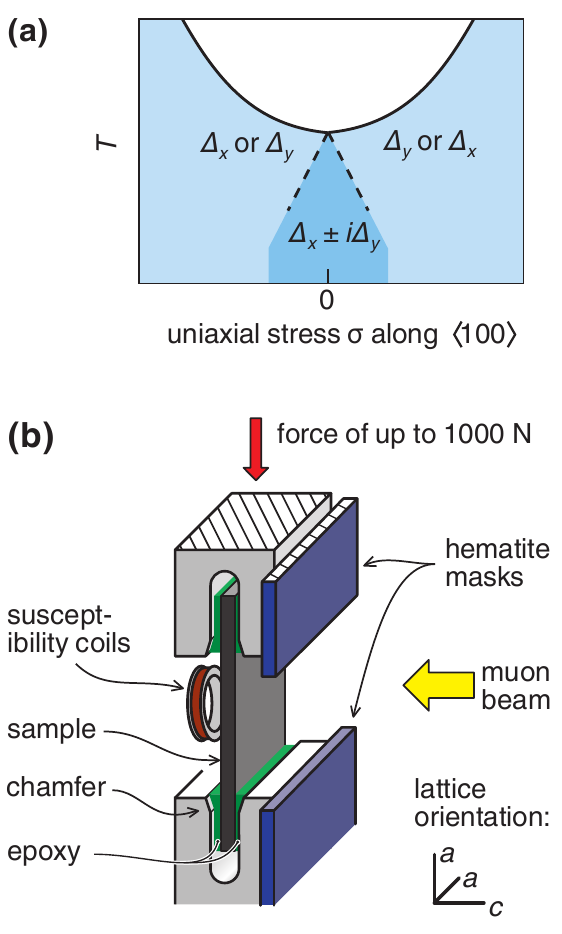}
\caption{\label{fig1}\textbf{Hypothesis.} \textbf{(a)} Schematic uniaxial stress-temperature phase diagram for
chiral superconductivity in defect-free \SRO{}; ``$\Delta_x$'' stands for either $p_x$ or $d_{xz}$. In the
mean field approximation, $T_\text{c}$ and $T_\text{TRSB}$ for a chiral order split linearly with applied stress $\sigma$ in
the limit $\sigma \rightarrow 0$. We indicate a quadratic upturn in $T_\text{c}$ at higher stresses, as seen in
measurements, and do not speculate how $T_\text{TRSB}$ should evolve at large $|\sigma|$. \textbf{(b)} Schematic of the
sample setup for $\mu$SR. Hematite masks screen the portions of the holder in the beam.  The chamfers are intended to
smooth the stress profile and reduce shear stresses in the sample.}
\end{figure}

\section{Results on unstressed \SRO{}}

We begin by presenting the results from three samples (labelled A-C) at zero stress.
%We begin presentation of results with those from %three samples (labelled A-C) at zero stress. 
\SRO{} is grown using a floating zone technique, and growth proceeds along an in-plane lattice
direction~\cite{Bobowski19}. For Sample A this was approximately a $\langle 110 \rangle$ direction, and for
Samples B and C a $\langle 100 \rangle$ direction.  Samples A and C have critical temperatures of 1.38 and
1.35~K, respectively: close to the clean-sample limit of 1.50~K~\cite{Mackenzie98}, while Sample B has
$T_\text{c} = 1.22$~K, indicating a larger number of defects. All $T_\text{c}$'s of unstressed \SRO{} were
determined by heat capacity and/or transverse-field $\mu$SR measurements, both of which are bulk-sensitive
probes. 

\begin{figure}
\includegraphics[width=57mm]{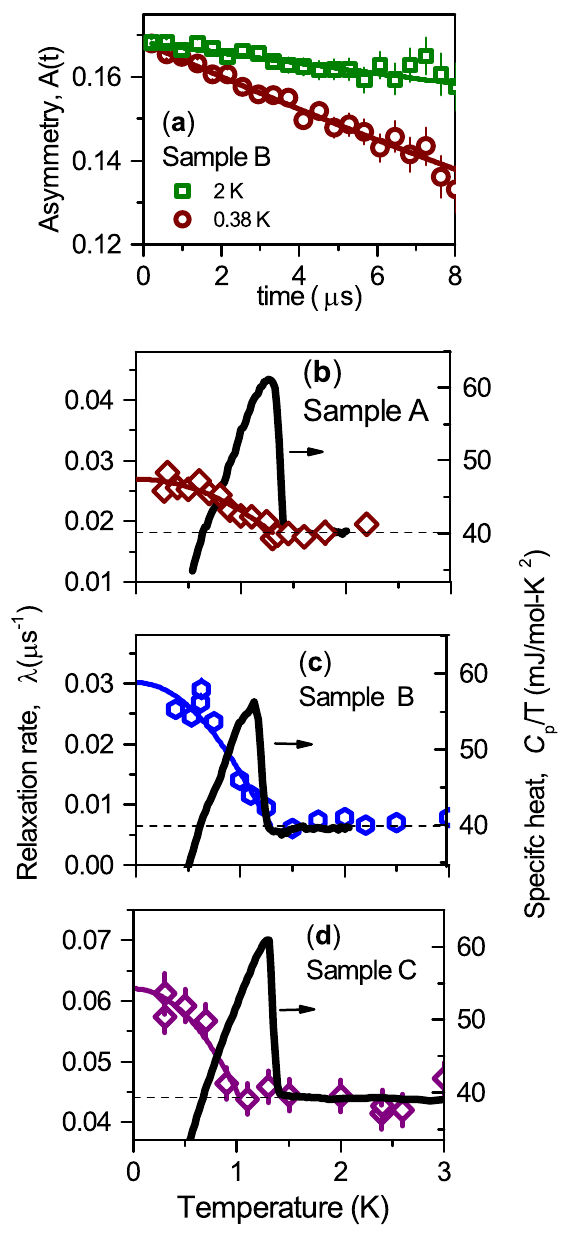}
\caption{\label{fig2} \textbf{FIGURE:} \textbf{(a)} Example of zero-field $\mu$SR asymmetry time spectra $A(t)$ of
Sample B above and below $T_\text{TRSB}$.  \textbf{(b - d)}Comparison of the temperature dependence of the zero field
muon relaxation rate $\lambda(T)$  (left scale) and heat capacity data (right scale) for Samples A, B and C, all under
zero stress. To determine $T_\text{TRSB}$, $\lambda(T)$ is fit with a quadratic form: $\lambda(T) = \lambda_0 + a[1 -
(T/T_\text{TRSB})^2]$ for $T<T_\text{TRSB}$, and $\lambda(T) = \lambda_0$ for $T>T_\text{TRSB}$.} 
\end{figure}

Measurements were performed with 4.2~MeV muons, which penetrate to a depth of $\sim$0.1~mm, using the
Dolly instrument on the $\pi$E1 beamline at the Paul Scherrer Institute. Decay positrons are counted by two
detectors placed forward and backward along the incident beam direction. The asymmetry $A(t)$ in the count
rate between them is proportional to the average muon spin polarization at time $t$. We employ a method to
extract $A(t)$ from the raw count rates known as single histogram fitting (explained in the supplementary
material) that reduces sensitivity to drifts in the instrumentation. 

Zero-field $A(t)$ data for Sample A at two temperatures are shown in Fig.~\ref{fig2}. Slow muon spin
relaxation, due mainly to nuclear magnetic fields, is observed at the higher $T$, and faster relaxation at the
lower $T$, indicating onset of an additional relaxation process. External fields are compensated to better
than 1~$\mu$T, so this behavior is not a consequence of appearance of vortices. The additional relaxation has
an exponential form, meaning that $A(T<T_\text{TRSB})/A(T>T_\text{TRSB})$ is approximately exponential. For a
static relaxation process this quantity is a Fourier transform of the internal field distribution, and it
indicates a broad field distribution characteristic of fields from dilute sources~\cite{Luke98}.

The muon spin relaxation rate $\lambda$ at each temperature is obtained by fitting the following model:
\[
A(T, t) = A_\text{sam} e^{-\lambda(T) t} + A_\text{bkg}.
\]
$A_\text{bkg}$ is a background constant to account for muons that implant into non-superconducting material
such as cryostat walls, and $A_\text{sam}$ is the sample signal strength. For all samples in this paper,
$A_\text{bkg}$ and $A_\text{sam}$ are determined from transverse-field $\mu$SR measurements at a
temperature well below $T_\text{c}$, leaving $\lambda$ as the sole fitting parameter.  In a field applied
transverse to the muon polarization, muons in nonsuperconducting, nonmagnetic material precess in-phase, while
precession in superconducting material quickly decoheres due to the field inhomogeneity introduced by the
vortex lattice. These contrasting behaviours allow the two muon populations to be distinguished.  

As shown in Fig. \ref{fig2}, a phenomenological fit to $\lambda(T)$  yields $T_\text{TRSB} = 1.30 \pm 0.06$~K
for Sample A and $1.3 \pm 0.1$~K for Sample B: the same within error bars, despite Sample B's lower
$T_\text{c}$. (All error bars in this paper are one standard deviation.) However susceptibility measurements
reveal that $T_\text{c}$ of Sample B is not homogeneous: the transition in susceptibility is broadened and
$0.25$~K above the transition seen in heat capacity, likely due to internal strains that locally induce higher
$T_\text{c}$~\cite{Maeno98, Steppke17}. For Sample C, in contrast, $T_\text{TRSB}$ is $1.03 \pm 0.08$~K, which
is below its $T_\text{c}$. 

These measurements on unstressed \SRO{} provide two important results. (1) The onset of enhanced muon spin relaxation is
sharp; it is a transition rather than a crossover.  Previously-published data sets \cite{Luke98, Luke00, Shiroka12,
Higemoto14} do not all have enough points above $T_\text{TRSB}$ to resolve this. It is an important point because there
is a known mechanism that can give weak exponential muon spin relaxation: fluctuations of weak ferromagnetism, as seen in YbNi$_4$P$_2$~\cite{Spehling12} and CeFePO~\cite{Lausberg12}.  The distinguishing feature is that in this case
the relaxation enhancement fades gradually over an order-of-magnitude increase in temperature, rather than at a
transition. (2) In combination with Sample D (shown below), we observe $T_\text{TRSB}$ to be within $\sim$0.1~K of
$T_\text{c}$ for three samples, and to be suppressed relative to $T_\text{c}$ for one. In combination with published
results~\cite{Luke98, Luke00, Shiroka12, Higemoto14}, the phenomenology appears to be that $T_\text{TRSB}$ can be
suppressed below $T_\text{c}$, but cannot exceed $T_\text{c}$. In attempting to resolve contradictory measurements on
\SRO{} it has been asked whether the enhanced muon spin relaxation is in fact related to the superconductivity.  The
combination of transition-like onset and correlation of $T_\text{TRSB}$ with $T_\text{c}$ is strong evidence that it is.

\section{Results under uniaxial stress}

Stress $\sigma$ is applied along a $\langle 100 \rangle$ lattice direction, which couples strongly to the electronic
structure~\cite{Hicks14Science}. At $\sigma = -0.7$~GPa (where $\sigma < 0$ denotes compression) there is a Fermi
surface topological transition (a Lifshitz transition), at which $T_\text{c}$ peaks at 3.5~K~\cite{Steppke17,
Barber19VHS}. In the limit $\sigma \rightarrow 0$ and in the mean field approximation, the ratio of the slopes
$|dT_\text{TRSB}/d\sigma|$ and $|dT_\text{c}/d\sigma|$ is inverse to the ratios of the associated heat capacity jumps
$\Delta C/T$. The experimental upper limit on the heat capacity jump at any second transition is $\sim$5\% of that
at the superconducting transition~\cite{Li19}, and therefore in the hypothesis figure [Fig.~\ref{fig1} (a)] we
illustrate a steep initial $|dT_\text{TRSB}/d\sigma|$.  The physical meaning  is that under the hypothesis of chiral
order, the energy difference between nonchiral and chiral superconductivity, for example $p_x$ versus $p_x \pm ip_y$,
must be small in unstressed \SRO{}. In support of the plausibility of this circumstance, we note that weak-coupling
calculations such as in Ref.~\cite{Roising19} give nodes that have very narrow opening angles. The condensation energy
gained from filling in such nodes by introducing chirality would be small.

Uniaxial stress was applied using a piezoelectric-based cell adapted to the sample size requirements for $\mu$SR,
described in detail in Ref.~\cite{Cell18}. The maximum force this cell can apply is 1000~N.  The beam is $\sim$1~cm in
diameter, and for a decent count rate the sample area facing the beam should be at least $\sim$10~mm$^2$.  A schematic
of a mounted sample is shown in Fig.~\ref{fig1}(b): the sample is a plate that is thick enough both to stop the muons and
to resist buckling at the highest applied stress.  Hematite masks screen the portions of the sample holder exposed to
the beam; the strong antiferromagnetism of hematite relaxes the polarization of muons that implant into these masks
within $\sim 10$~ ns, allowing these muons to be excluded from analysis. Being antiferromagnetic, the masks do not
generate long-range stray fields. Each sample holder incorporates a force sensor based on strain gauges.  Also, for most measurements a
pair of concentric coils was placed behind the sample, for in situ measurement of $T_\text{c}$ through the diamagnetic
shielding of the superconducting state. Measured values of $T_\text{c}$ were used to calibrate the force sensors
following the stress dependence reported in Ref.~\cite{Barber19VHS}.  Three samples, labelled D--F, were tested under
uniaxial stress. Samples E and F were cut from Sample C. 

\begin{figure}
\includegraphics[width=114mm]{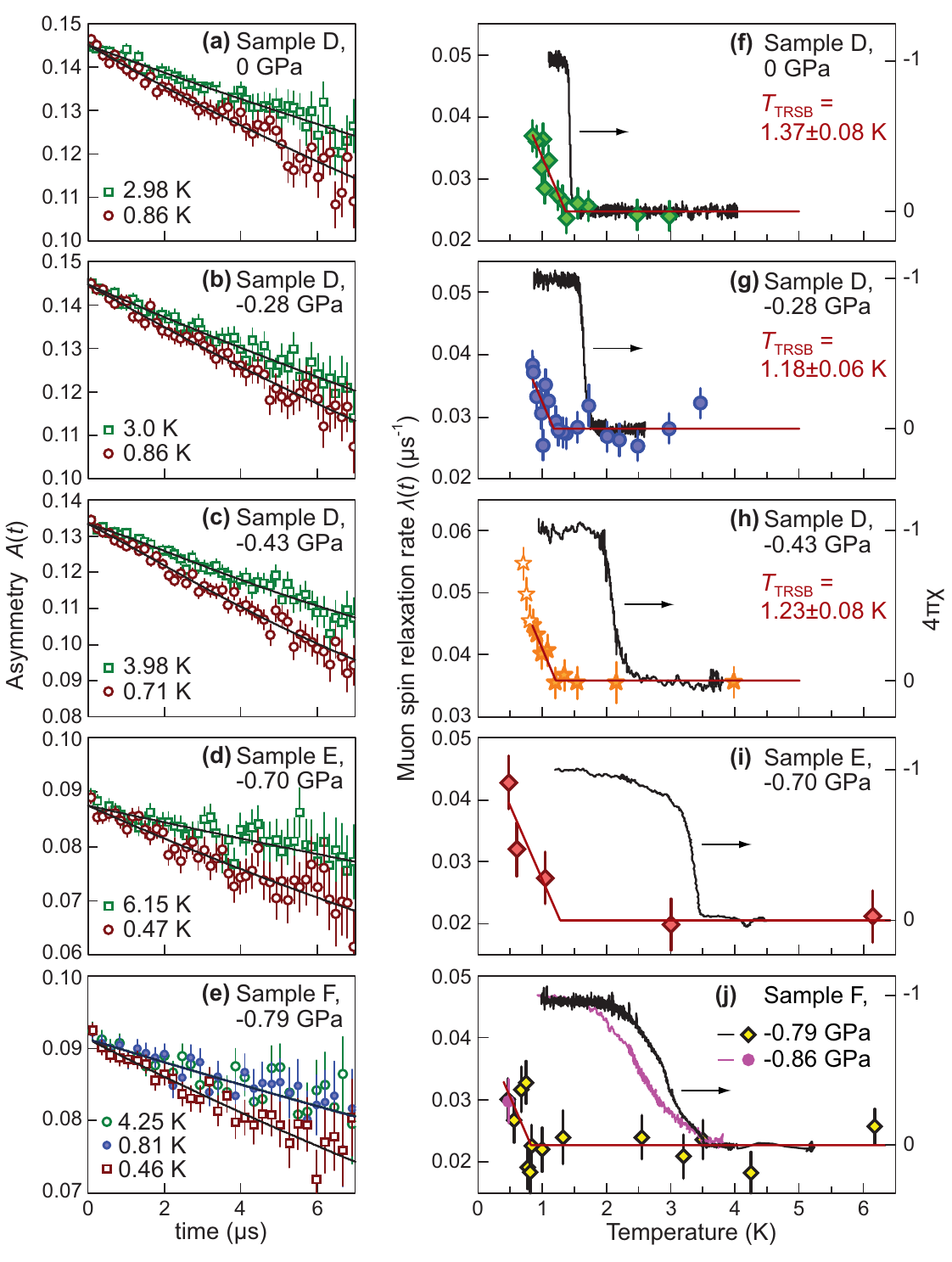}
\caption{\label{fig3}Left-hand panels: Zero-field $\mu$SR asymmetry $A(t)$ at a high temperature and at the lowest
temperature reached. \textbf{(a--c)} Sample D at 0~GPa,  -0.28~GPa and  -0.43~GPa, \textbf{(d)} Sample E at -0.70~GPa,
and \textbf{(e)} Sample F at -0.79~GPa plus one data point at -0.86~GPa. (Negative values denote compression.)
\textbf{(f--j)}: Temperature dependence of the muon spin relaxation rate $\lambda$, and in situ diamagnetic
susceptibility data for the samples and stresses of the left-hand panels. The applied field for the susceptibility
measurements was $\sim 10$~$\mu$T. Heat capacity and transverse-field $\mu$SR data show that the samples are fully
superconducting, so we identify the extrema of the susceptibility signal as $4\pi\chi = 0$ and $-1$. The fits to
$\lambda(T)$ (red lines) are explained in the text. Note that for panels (f--h), which are all on Sample D, to avoid
biasing the fit the fitting range is the same in each panel, which excludes the three open points in panel (h).}
\end{figure}

Results are shown in Fig.~3. Sample D, with the zero-stress $T_{\rm c}$ = 1.39 K, was measured at 0, -0.28, and
-0.43~GPa.  $A_\text{sam}$ and $A_\text{bkg}$ were determined independently
at each stress. The relaxation enhancement remains exponential at each stress, and it can be seen in panels (f--h) that
although $T_\text{c}$ increases under the applied stress, $T_\text{TRSB}$ remains low.  Because the data here do not
extend to very low temperature (due to the large mass and poor thermal conductance of the pressure apparatus), a
simpler, linear form is used to fit $\lambda(T)$ and extract $T_\text{TRSB}$:
\[
\lambda = \left\{
	\begin{array}{ll}
	\lambda_0 + b \times ( T_\text{TRSB} - T), & T < T_\text{TRSB} \\
	\lambda_0, & T > T_\text{TRSB} \\
	\end{array}
	\right.
\]
The slope $b$ is a common fitting parameter among all three stresses, while $T_\text{TRSB}$ and $\lambda_0$ are obtained
independently at each stress. This fit gives $T_\text{TRSB} = 1.37 \pm 0.08$~K at 0~GPa, $1.18 \pm 0.06$~K at -0.28~GPa,
and $1.23 \pm 0.08$~K at -0.43~GPa.  Although small, a stress dependence of $T_\text{TRSB}$ is resolved: the
probability that $T_\text{TRSB}$ is lower at -0.28 GPa than at 0 GPa is 98\%.

\begin{figure}
\includegraphics[width=57mm]{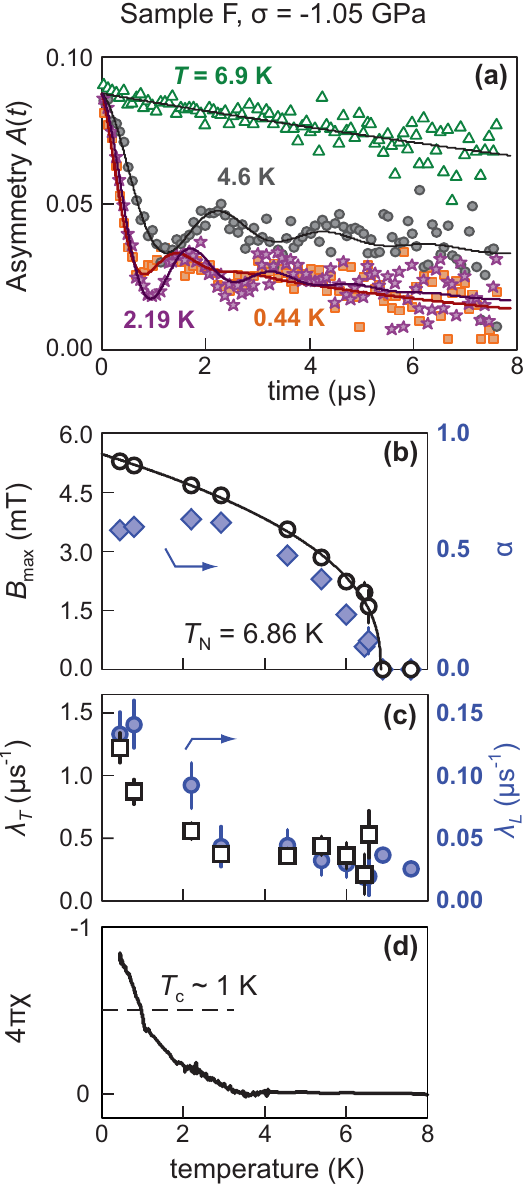}
\caption{\label{fig4}\textbf{Magnetic order.} \textbf{(a)} Zero-field asymmetry $A(t)$ at various temperatures for
Sample F at -1.05~GPa. \textbf{(b)} The maximum internal field $B_\text{max}$ and transverse signal fraction $\alpha$ as
a function of temperature. The fit to $B_\text{max}$ gives $T_N = 6.86$~K.
\textbf{(c)} Transverse and longitudinal relaxation rates versus temperature. Error bars, when not shown, are smaller
than the symbol. \textbf{(d)} In situ diamagnetic susceptibility data.}
\end{figure}

Samples E and F had smaller total cross-sections than Sample D, allowing higher stresses to be reached. Sample
E was measured at -0.70~GPa, right at the peak in $T_\text{c}$, and Sample F slightly beyond, at -0.79~GPa.
Adopting the same phenomenological fit as applied to Sample D, with the same slope $b$, $T_\text{TRSB}$ of
Sample F at -0.79~GPa is determined to be $0.82\pm0.09$~K. Although this is a low value, this sample was
extracted from sample C, whose zero-stress $T_\text{TRSB}$ was $1.03 \pm 0.08$~K, and we cannot firmly
conclude that $T_\text{TRSB}$ was suppressed by the stress. It is however clear that even when $T_\text{c}$ is
at its maximum, $T_\text{TRSB}$ is low. A single data point at a yet higher stress, -0.86~GPa, indicates that
the time-reversal symmetry breaking is still present.

Going further, bulk magnetic order appeared in Sample F at $\sim$ -1.0~GPa. $A(t)$ at various temperatures is shown in
Fig.~4(a): the oscillations are an unmistakable indication of long-range magnetic order. We fit the following form to
$A(t)$:
\[
A(t) = \alpha \, j_0(2\pi\gamma_\mu B_\text{max} t )e^{-\lambda_\text{T}t} + (1-\alpha) \, e^{-\lambda_\text{L}t}.
\]
Here, $j_0$ is a zeroth-order Bessel function, which is the Fourier transform of the Overhauser field distribution,
$p(B) = 2/\pi\sqrt{B_\text{max}^2 - B^2}$, expected for an incommensurate spin density wave. A damped cosine form,
expected for commensurate magnetic order or ferromagnetism, does not fit well. $B_\text{max}$ is the magnetic hyperfine
field at the peaks of the SDW, and $\gamma_\mu$ the muon gyromagnetic ratio. $\alpha$ is the oscillating signal fraction
due to muons experiencing magnetic hyperfine fields transverse to the initial muon spin polarization. Since the magnetic
order can  also generate longitudinal field components at individual muon sites, which do not cause muon spin
precession, $\alpha$ is a lower bound for the magnetic volume fraction.  $\lambda_\text{T}$ and $\lambda_\text{L}$
describe an additional static line broadening and a slow dynamical spin relaxation, respectively.  Results of fitting
are shown in Fig.~4.  $\alpha$ saturates at $\approx$60 \%, and $B_\text{max} (T \rightarrow 0)$ is $5.5 \pm 0.1$~mT.
Fitting $B_\text{max}(T)$ gives a N\'{e}el temperature $T_N$ of 6.86~K. $\lambda_T$ and $\lambda_L$ strongly increase
below $\approx$2~K. The effect can be seen directly in Fig.~4(a): more oscillations are resolvable at 2.19 than at
0.44~K. Susceptibility data [Fig.~4(d)] show that the sample is superconducting with $T_\text{c} \sim 1$~K (though with
a broad transition), so we conclude that the increase in $\lambda_\text{T}$ and $\lambda_\text{L}$ at low
temperatures is a consequence of microscopic coexistence of the superconductivity and magnetism. As further evidence
that the magnetic order is a spin density wave, we note that there is no anomaly in the susceptibility data at $T \sim
7$~K that would indicate ferromagnetism.

In unstressed \SRO{}, inelastic neutron scattering reveals strong magnetic fluctuations along columns in momentum space
$q = (\pm 0.3, \pm 0.3, q_z)$, due to nesting between the $\alpha$ and $\beta$ Fermi surfaces~\cite{Sidis99}.
Substitution of a few per cent Ti on the Ru site induces static order at this $\mathbf{q}$, and at 9\% Ti the $T
\rightarrow 0$ ordered moment is 0.3~$\mu_B$/Ru~\cite{Minakata01, Braden02Ti}. Although the electronic structure of
\SRO{} also introduces susceptibilities at other $\mathbf{q}$'s~\cite{Steffens19, Cobo16}, it is a reasonable hypothesis
that the stress-induced and Ti-induced magnetic orders are related. Sr$_2$Ru$_{0.91}$Ti$_{0.09}$O$_4$ has been studied
with muons~\cite{Carlo12}: it has $T_N \approx 20$~K, and the first minimum in $A(t)$ occurs at $\approx$0.2~$\mu$s,
against $\approx$1~$\mu$s for the stress-induced magnetic order here.  This observation suggests an ordered moment for
the stress-induced order of $\sim$0.06~$\mu_B$/Ru.

In functional renormalization group calculations reported in Ref.~\cite{Liu17}, uniaxial stress was predicted
to induce the formation of SDW order \textit{before} reaching the Lifshitz transition.  However, here it is
clear that the magnetic order onsets beyond the Lifshitz transition: the -0.86~GPa data point of Sample F
falls between the Lifshitz transition and the onset of SDW order. Our data are summarized by the phase diagram
in Fig.~5.

\section{Discussion}

We focus our Discussion on the observation for $|\sigma| < 1.0$~GPa of stress-induced splitting between $T_\text{c}$ and
$T_\text{TRSB}$. Splitting between $T_\text{c}$ and $T_\text{TRSB}$ has been observed previously in a few materials, but
not with the clarity attained here. In UPt$_3$, a splitting of $\sim$0.05~K was observed~\cite{Luke93}, although
enhanced muon spin relaxation was not seen at all in a later report~\cite{Reotier95}. In both
Ba$_{1-x}$K$_{x}$Fe$_2$As$_2$ and Pr$_{1-x}$La$_x$Pt$_4$Ge$_{12}$ there is a potential splitting of a few
K~\cite{Grinenko17, Zhang19}, but resolution in both cases is limited by the transition widths and small scale of the
increase in $\lambda$.

\begin{figure}
\includegraphics[width=100mm]{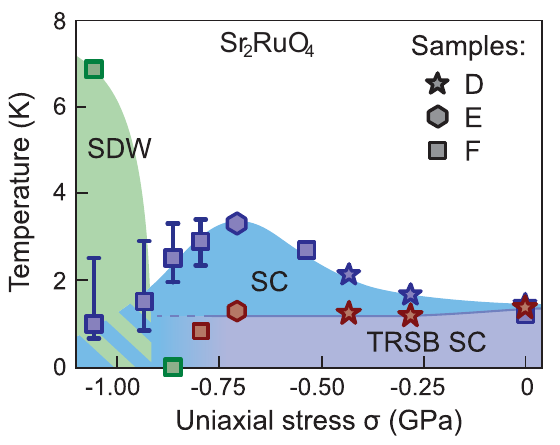}
\caption{\label{fig5}Stress-temperature phase diagram of \SRO{} based on the data presented here. For conversion to
strain, the low-temperature Young's modulus for compression along a $\langle 100 \rangle$ direction is
160~GPa~\cite{Barber19VHS}.}
\end{figure}

The splitting between $T_\text{c}$ and $T_\text{TRSB}$ rules out the possibility that enhanced
muon spin relaxation is an artefact of a conventional superconducting transition. This is an especially
important observation in the absence of a known microscopic mechanism that reconciles the different field
scales observed in $\mu$SR and scanning SQUID magnetometry measurements. It rules out, for example, that the
enhanced relaxation is a consequence of compression of flux from magnetic inclusions, present in all large
\SRO{} samples, due to Meissner screening. It rules out any mechanism based on
interaction of magnetic fluctuations and conventional superconductivity.

There is no known magnetic mechanism that could account for the $\mu$SR data. We have noted  that
relaxation by weak ferromagnetic fluctuations is not consistent with the observed transition-like onset of enhanced muon
spin depolarization. A glassy magnetic state could reproduce the broad distribution of fields implied by exponential
relaxation, however even dilute spin glasses typically give two orders of magnitude stronger relaxation~\cite{Wu94}.
Finally, we have observed magnetic order in clean \SRO{} at high stress, and its qualitative appearance in $\mu$SR data
is completely different.

We conclude, from the correlation between $T_\text{c}$ and $T_\text{TRSB}$ in unstressed \SRO{} and the absence of known
magnetic mechanisms, that the enhanced muon spin relaxation is a property of the superconductivity.  The fact that
$T_\text{TRSB}$ can split from $T_\text{c}$ shows further that it is a transition of the superconducting state. This
provides strong support for the hypothesis, widely accepted  but not  rigorously proved, that enhanced muon spin
relaxation is a product of TRSB superconductivity. The observed stress-induced splitting furthermore follows qualitative
expectations for chiral superconductivity in \SRO{}. We note that recent ultrasound data~\cite{Ghosh20} also indicate
two-component superconductivity.

A uniaxial stress of -0.28~GPa was observed to suppress $T_\text{TRSB}$ by $\sim$0.2~K. Beyond this stress,
$T_\text{TRSB}$ appears not to evolve strongly, an observation that superficially contrasts with the prediction from
Landau theory that uniaxial stress should suppress $T_\text{TRSB}$. However approaching the Lifshitz transition is a
strongly nonlinear process, and linear extrapolations based on Landau theory are unlikely to be valid at large
$|\sigma|$. Stress-driven suppression of $T_\text{TRSB}$ could be balanced by the overall strengthening of
superconductivity. At low stresses, we note that disorder~\cite{Yu19, Hicks14Science} and fluctuations~\cite{Fischer16}
are predicted to round off cusps, potentially weaking the observed stress dependence of $T_\text{TRSB}$. Under the
hypothesis of chirality, we speculate further that disorder could be the mechanism that splits $T_\text{TRSB}$ and
$T_\text{c}$ in some samples even at zero stress. Dislocations and inclusions are common in large samples of \SRO{}, and
the growth direction of the sample could give them a preferred orientation, lifting the tetragonal symmetry of the
system.

Reconciling evidence for even-parity pairing with chirality leads to the consideration of an order parameter in the
$E_g$ representation: $d_{xz} \pm id_{yz}$. When orbital degrees of freedom are neglected it is an unexpected order
parameter because the line node at $k_z=0$ implies interlayer pairing in a metal with low $k_z$ dispersion; its
$T_\text{c}$ is strongly suppressed in the weak-coupling calculations of Ref.~\cite{Roising19}. To get around this
difficulty, alternative TRSB order parameters without horizontal line nodes have been proposed: $d \pm is$ and
$d_{x^2-y^2} \pm i g_{xy(x^2-y^2)}$. Although they depend on accidental degeneracy to obtain $T_\text{TRSB} \approx
T_\text{c}$ on a tetragonal lattice, there is theoretical support that this 
may be realized in
\SRO{}~\cite{Romer19, Kivelson20}. On the other hand, including orbital degrees of freedom allows the possibility of
interorbital pairing driven by on-site Hund's rule coupling. This mechanism becomes feasible when
spin-orbit and Hund's-rule couplings are non-negligible in comparison with the Fermi energy; how strong they must be is
a subject of debate~\cite{Puetter12, Hoshino15, Ramires19, Suh19}. As yet, there are no widely-accepted examples of
this type of superconductivity, however it allows  that $E_g$ symmetry is encoded in the local orbital
degrees of freedom rather than the $k$ dependence of the gap, such that interlayer pairing is no longer required. In
Ref.~\cite{Suh19} it is proposed that introducing momentum dependence to the spin-orbit coupling can favour
interorbital $E_g$ order in \SRO{}.

In conclusion, the data presented here are consistent with chiral superconductivity in \SRO{}. This is not just an
unusual phase, but imposes strong qualitative constraints on models of pairing, possibly requiring a new mechanism.
The superconductivity of \SRO{} remains an important question, and we encourage further exploration.

\textbf{Acknowledgement.} This work has been supported financially by the Deutsche Forschungsgemeinschaft (GR 4667/1,
GRK 1621, and SFB 1143 project C02) and the Max Planck Society. YM, TM, and JB acknowledge the financial support of JSPS
Kakenhi (JP15H5852 and JP15K21717) and the JSPS Core-to-Core Program. NK acknowledges the financial support from JSPS
Kakenhi (No. JP18K04715) and JST-Mirai Program (No. JPMJMI18A3). AN acknowledges funding from the European Union's
Horizon 2020 research and innovation program under the Marie Sklodowska-Curie grant agreement No 701647. This work was
performed partially at the Swiss Muon Source (S$\mu$S), PSI, Villigen. We acknowledge fruitful discussions with A.
Amato, E. Babaev, S. Blundell, A. Charnukha, D. Efremov, I. Eremin, C.  Kallin, A. Ramires, B. Ramshaw, T. Scaffidi, C.
Timm and S. Yonezawa. We also acknowledge H.-S. Xu for his contribution in the crystal growth as well as  T. Shiroka and
C. Wang for technical support. We thank A. Gilman, P. P. Orth, and R. M. Fernandes for results from Landau theory of a
two-component order parameter.
\\ 
\\

\renewcommand{\theequation}{S\arabic{equation}}
\renewcommand{\thefigure}{S\arabic{figure}}
\renewcommand{\thetable}{S\arabic{table}}
\setcounter{equation}{0}
\setcounter{figure}{0}
\setcounter{table}{0}

\section{Supplementary material: Split superconducting and time-reversal symmetry-breaking transitions, and magnetic order in \SRO{} under uniaxial stress}

\subsection{Methods}

Single crystals of \SRO{} were grown by a floating zone method~\cite{Bobowski19}. With the exception of Sample
C, all samples studied here were either cleaved or ground into plates, exposing the interior of the as-grown
rod to the muon beam; this is technically relevant information because due to differential evaporation of Ru
and Sr during the growth, the interior tends to have a higher density of inclusion phases, especially Ru,
SrRuO$_3$, and Sr$_3$Ru$_2$O$_7$. Crystals grow along an in-plane direction, and to obtain samples of
sufficient length for measurement under uniaxial stress we were obliged to select samples that happened to
grow nearly along a $\langle 100 \rangle$ direction. Samples were mounted into holders as shown in Fig.~\ref{FigS1}
using Stycast 2850 epoxy; the epoxy layers were generally 50--100~$\mu$m thick. For Samples E and F, three
additional steps were taken to improve the chances of reaching high stresses without fracturing the sample.
(1) They were cut at a $\sim10^\circ$ angle with respect to the $ab$ plane, so that shear stresses in the
sample do not align with cleave planes.  (2) 10~$\mu$m-thick titanium foils were affixed to their surfaces
with Stycast 1266. (3) The slots in the holder were chamfered, as shown in Fig.~1(b), to smooth the interface
between the free and clamped portions of the sample.

In the $\mu$SR system, an upstream detector triggers a timer when a muon enters the system. Muons entering the
system meet one of four fates. (1) Passage through the holder without implantation.  These muons are detected
by a downstream veto detector, and positron counts from their decay are rejected. (2) Implantation in the
hematite masks. [See Fig.~1(b).] These muons depolarize very rapidly.  (3) Implantation in material other than
the sample or hematite; this gives the background asymmetry $A_\text{bkg}$.  (4) Implantation in the sample.

We now explain the single histogram analysis. Muon asymmetry is typically obtained through direct comparison
of the count rates in two detectors. In the single-histogram analysis, data from each detector are analysed
separately. This reduces sensitivity to instrumentation drifts while increasing sensitivity to background; it
improves precision in determination of relative values of $\lambda$ while increasing uncertainty in the
absolute values.

Upon decay, emitted positrons are detected by either a forward or backward detector, and counts are binned by
time after the trigger. In principle the asymmetry in the count rate, $A(t) = [N_1(t) - N_2(t)]/[N_1(t) +
N_2(t)]$ where $N_i(t)$ is the number of counts in detector $i$ at time $t$, is proportional to the average
muon polarisation at time $t$. In practice, each detector has a ``dark" count rate, $N_{\mathrm{dark},i}$, that
must first be subtracted from $N_i(t)$.  Typically, dark rates are measured in situ, but when the relaxation is slow (as in \SRO{})
fitted relaxation rates become sensitive to errors in the dark rates. In the single histogram analysis,
the $N_{\text{dark},i}$ are instead treated as fitting parameters. The counts $N_i(t)$ are fit by 
\[
N_i(t) = N_{0,i}[1 + A_i(t)]e^{-t/\tau_\mu} + N_{\mathrm{dark},i},
\]
where $\tau_\mu$ is the muon lifetime and $N_{0,i}$ is a fitting parameter setting the overall number of
synchronous (\textit{i.e.} not dark) counts. $A_i(t)$ is a hypothesised functional form for the emission asymmetry; it has the same form
for both detectors, though may have detector-dependent parameters (as indexed by the subscript $i$). We take
$A_i(t) = A_{\mathrm{sam},i} e^{-\lambda t} + A_{\mathrm{bkg},i}$.  $\lambda$, the relaxation rate, is
determined from fits to both detector signals simultaneously, using the MUSRFIT
program~\cite{Suter12}. In this study we generally plot the combined asymmetry $A(t)$ as $A(t) \equiv \frac{1}{2}[A_1(t) - A_2(t)]$, obtaining the fitted
form of $A_i(t)$ as just described and taking the experimental $A_i(t)$ as: 
\[ 
A_i(t) = \frac{N_i(t) - N_{\mathrm{dark},i}}{N_{0,i}\exp(-t/\tau_\mu)} - 1.  
\]
 $A_{\mathrm{sam},i}$ and $A_{\mathrm{bkg},i}$ need to be determined separately from
transverse-field measurements, because when relaxation is slow the single histogram fits have low
resolving power between these parameters and $N_{\text{dark},i}$.

For simplicity, a non-relaxing background $A_\text{bkg}$ was assumed for all samples.  This is not strictly correct. For
example, between -0.28 and -0.43~GPa a chip broke from Sample D, exposing some brass material of the ac susceptibility
coil holder, which is relaxing due to the nuclear magnetic moments of Cu. If instead $A_\text{bkg}$ is taken to relax
following an assumed functional form, the single histogram analysis gives different absolute values for $\lambda$. This
is the reason why the fit procedure introduces an uncertainty on the absolute value of $\lambda$. We show below that
assuming different backgrounds has no substantial effect on temperature-dependent changes in $\lambda$.

Although here we fit zero-field $\mu$SR time spectra to simple exponential decays, it is also common to
analyse zero-field time spectra by $ A(t) = A_\mathrm{bkg} + A_\mathrm{sample} \times f_\text{GKT}(t) \times
e^{-\lambda t}$, where $f_\text{GKT}$ is the Gauss-Kubo-Toyabe function, $\mathrm{GKT} = \frac{1}{3} +
\frac{2}{3}(1-\sigma^2 t^2)\exp(-\frac{1}{2} \sigma^2 t^2)$. In principle, $f_\text{GKT}$ is the
contribution from dense sources such as nuclear dipoles, and $e^{-\lambda t}$ the contribution from dilute
sources. We tested a simpler phenomenological form on Sample D, $A(t) = A_\mathrm{bkg} + A_\mathrm{sam}
\exp(-\lambda t) \exp(-\frac{1}{2} \sigma^2 t^2)$, but found that the fit quality was not improved by the
added parameter.

\subsection {Effect of the background in the single histogram analysis}

\begin{figure}[ptb]
\includegraphics[width=120mm]{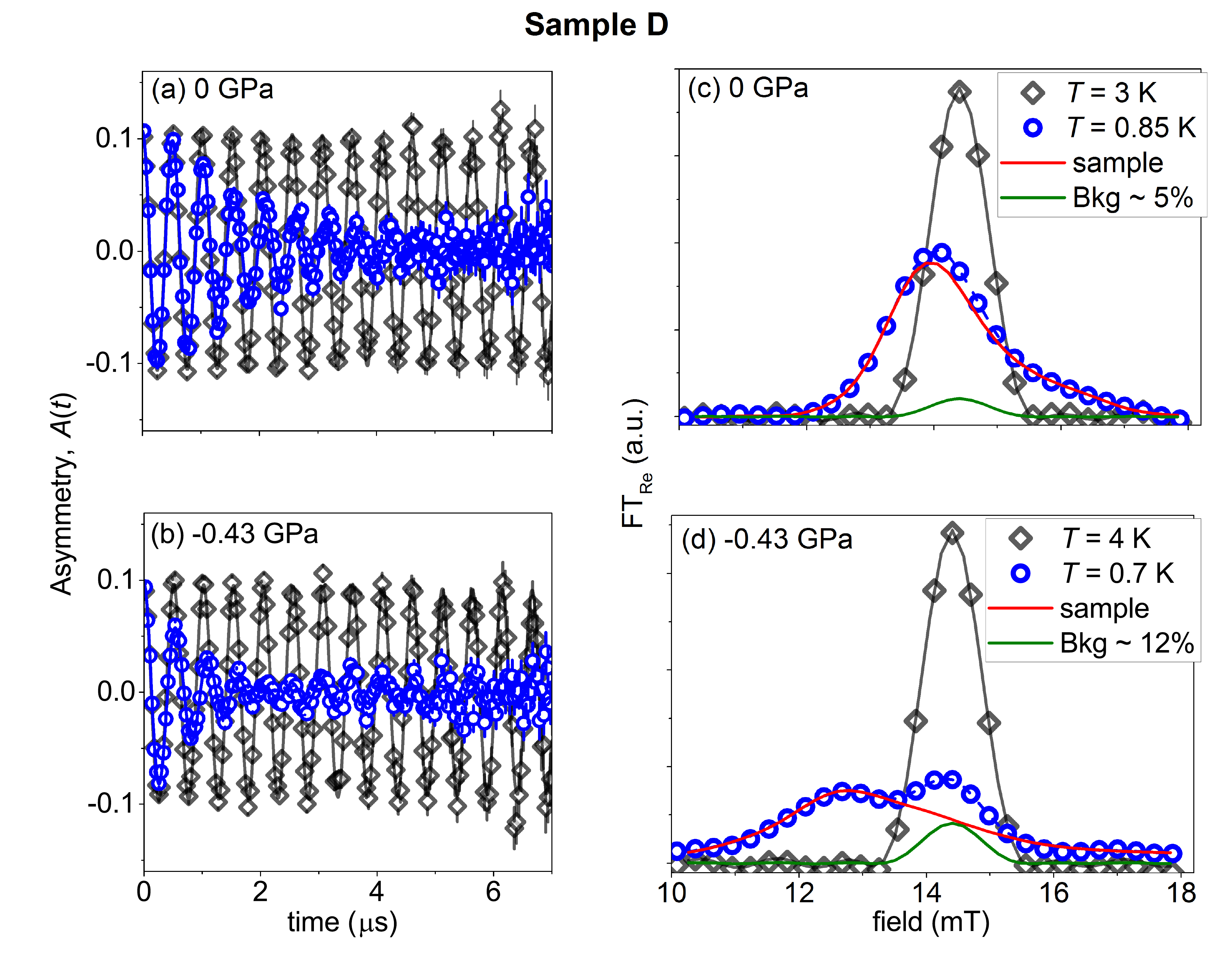}
\caption{\textbf{(a-b)} Transverse-field  $\mu$SR time spectra above and below $T_\mathrm{c}$, with a muon spin
polarization 45$^\circ$ with respect to the sample $c$ axis $\mathbf{\hat{c}}$ and $\mathbf{B}\parallel
\mathbf{\hat{c}}$. \textbf{(c-d)} Fourier transforms of the time spectra shown in panels (a) and (b). The fit to
determine the sample contribution is done in the time domain.} 
\label{TFdata2}
\end{figure}

Before each zero field measurements the ratio $A_\mathrm{bkg}/A_\mathrm{sam}$ was determined using
transverse field $\mu$SR. In Fig.~\ref{TFdata2} transverse-field data from Sample D at 0 and -0.43~GPa are
shown. Above $T_\mathrm{c}$ the muons precess without relaxation in the applied 14.5~mT field. The field was
applied along the $c$ axis, where $H_\text{c2}(T \rightarrow 0) \approx 70$~mT, and below $T_\mathrm{c}$
the field within the sample becomes highly inhomogeneous due to the vortex lattice.  Correspondingly, the
polarization of muons implanted in the sample relaxes rapidly.  At -0.43~GPa, it can be seen however that
there is residual oscillation that appears not to relax. This is due to the background muons; as
noted above, a chip broke from the sample between -0.28 and -0.43~GPa, exposing some of the susceptibility
coil holder (which was made of brass). To fit $A(t)$, we assume a field distribution that is a sum of a few
Gaussians, with one describing the background and the rest the vortex lattice~\cite{Maisuradze2009}. $A(t)$ is
fitted in the time domain to a Fourier transform of this assumed field distribution. Results are shown in
panels (c) and (d). By comparing the fitted amplitudes of the background and vortex contributions, we obtain
$A_\text{bkg}/A_\text{sam} \approx 0.12$ at -0.43~GPa, and 0.05 at 0 and -0.28~GPa. For the remaining samples,
$A_\text{bkg}/A_\text{sam}$ came to $\approx 0$ for Samples A, B, and F; $\approx 0.12$ for Sample C; and
$\approx 0.15$ for Sample E.

It was noted above that the single histogram analysis introduces uncertainty into absolute values of $\lambda$, but not
into temperature-dependent changes. We provide further details here. We show that taking a relaxing background, due to
possible magnetism in the background material, shifts the absolute values of $\lambda$ obtained from the single
histogram analysis, but does not substantially affect relative values.

\begin{figure}[ptb]
\includegraphics[width=160mm]{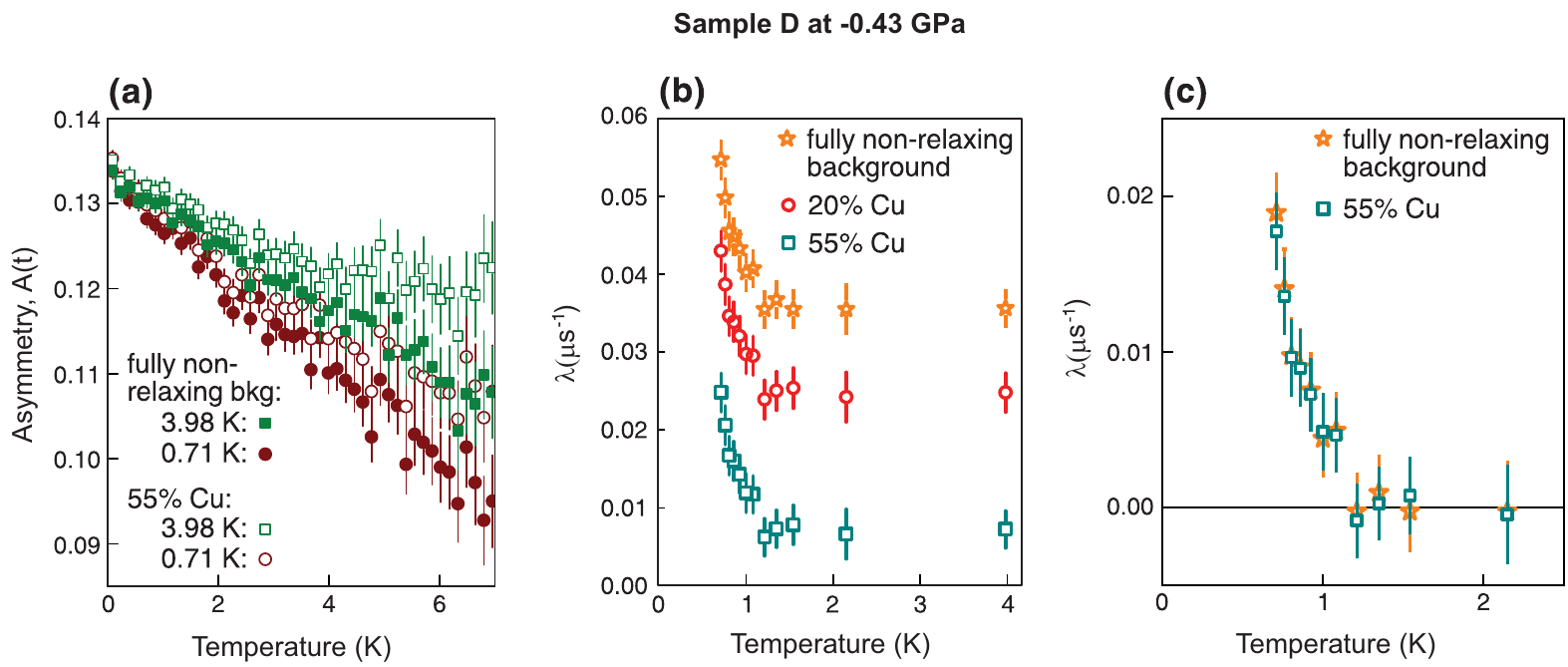}
\caption{\label{figS_bg}\textbf{Testing different backgrounds. (a)} $A(t)$ resulting from the single histogram analysis
of Sample D when a non-relaxing background term is assumed, and when a 55\% of the background is assumed to relax muon
polarization following, as expected for copper, the Kubo-Toyabe formula with $\sigma = 0.39$~$\mu$s$^{-1}$. \textbf{(b)}
Results for the relaxation rate $\lambda$ taking different portions of copper in the background. The dominant effect of
adding a relaxing element to the background is a uniform shift downward. \textbf{(c)} There is essentially no effect on
temperature-dependent relative changes in $\lambda$ extracted from the analysis.}
\end{figure}

The effect will have been particularly notable for Sample D at -0.43~GPa: the exposed brass is relaxing due to Cu
nuclear magnetic moments. The relaxation from copper can be accurately fit by the Gauss-Kubo-Toyabe form with $\sigma =
0.39$~$\mu$s$^{-1}$~\cite{Clawson83}. In Fig.~\ref{figS_bg}(a) we show results for $A(t)$ taking 20\% of the background
to relax the muon spins as copper does, and the remaining 80\% to be non-relaxing. This alters the fitted values of
$N_{\mathrm{dark},i}$, and so also the resulting $A(t)$.  Panel (b) shows $\lambda$ extracted from the single histogram
analysis with a non-relaxing background, this 20\% background, and also a 55\% background.  The 20\% copper case is seen
to reduce $\lambda$ for $T > T_\mathrm{TRSB}$ to a similar value as obtained for 0 and -0.28~GPa; because the background
was smaller for 0 and -0.28~GPa, these two measurements are more likely to yield correct absolute values.  The important
point  is revealed in panel (c): for all of these assumed backgrounds, the effect on temperature-dependent relative
values of $\lambda$ extracted from the single histogram analysis is negligible.

\subsection {Analysis of the ZF $\mu$SR data in the magnetic state}

\begin{figure}
	\includegraphics[width=80mm]{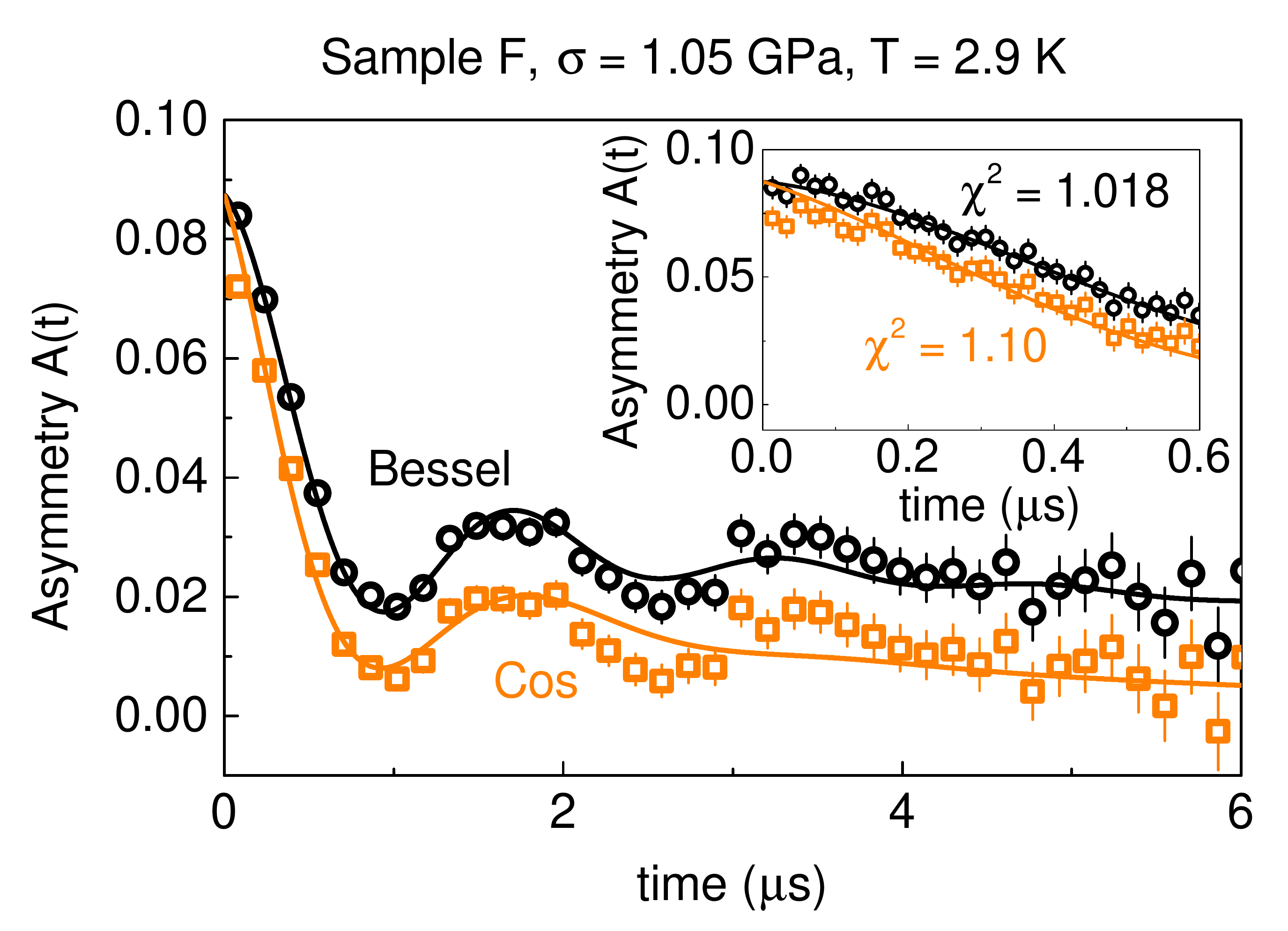}
	\caption{\label{Bessel_SM} 
	\textbf{Testing different models to describe magnetic order.} Zero-field asymmetry $A(t)$ in the
magnetic state for Sample F at -1.05~GPa, fitted with a Bessel function describing incommensurate spin
density wave order, and a damped cosine describing ferromagnetism or commensurate magnetic order. The Bessel
function is seen to give a better fit; the cosine fit does not capture oscillations after $\sim 2$~$\mu$s. Note that in
the single histogram analysis a functional form for $A(t)$ must be assumed to fit the dark count rate, which results in
separate data values for the two forms. Both are derived from a single data set.}
\end{figure}

To describe the internal field distribution in the magnetic state we used the Bessel function as discussed in the main
text. This field distribution is consistent with an incommensurate magnetic structure indicating a SDW state. It would
also be a reasonable hypothesis that tuning \SRO{} to peaks in the density of states induces ferromagnetism. Therefore,  we
show here that ferromagnetism does not give a good match to the data. The expected functional form for $A(t)$ for
ferromagnetism is a cosine:
\[
A(t) = \alpha \, \cos(2\pi\gamma_\mu B t )e^{-\lambda_\text{T}t} + (1-\alpha) \, e^{-\lambda_\text{L}t},
\]
where $B$ is now the average local magnetic field at the muon site induced by the magnetism, and other quantities are as
in the main text. This form also applies to commensurate magnetic order: reversing the field direction on every
other site would not alter $A(t)$. Results are shown in Fig.~\ref{Bessel_SM}. It is seen that the cosine fit does not
work as well: it does not reproduce the data at early time ($t \lesssim 0.1 \mu$s) or at $t > 2 \mu$s. In particular, it
does not capture the third peak in $A(t)$, at $t \approx 3.2$~$\mu$s, which is clearly present not only here (at
$T=2.9$~K), but also in the 2.19 and 4.4~K data in Fig.~4(a).

\subsection{Analysis of heat capacity jumps}

The hypothesized phase diagram, Fig.~1(a), indicates a slope $|dT_\text{TRSB}/d\sigma|$, where $\sigma$ is stress applied along a $\langle 100 \rangle$ lattice direction, that, for consistency with
heat capacity data under uniaxial stress~\cite{Li19}, far exceeds $|dT_\text{c}/d\sigma|$. Here, we explain further.

We take a general free energy
\[
F_0 = \frac{a}{2}\left(\Delta_x^2 + \Delta_y^2\right) + \frac{u}{4}\left(\Delta_x^4 + \Delta_y^4\right) + \frac{\beta}{2}\Delta_x^2\Delta_y^2 + \frac{\alpha}{2}\Delta_x^2\Delta_y^2\cos(2\theta) +
\frac{g \sigma}{2}\left(\Delta_x^2 - \Delta_y^2\right).
\]
$\Delta_x$ and $\Delta_y$ are respectively the amplitudes of the $x$ and $y$ components of the gap function, and $\theta$ is the phase difference between them. $\sigma$ is the applied uniaxial stress,
and $g$ a coupling constant. $a = a_0(T - T_{\text{c},0})$.

For a chiral state to occur, we require $\alpha > 0$ to ensure that the energy is minimized for $\theta = \pm \pi/2$, and $|\beta - \alpha| < u$ to ensure that both $\Delta_x$ and $\Delta_y$ become
nonzero below $T_{\text{c},0}$ in the unstrained sample. To obtain the mean-field phase diagram, this free energy is minimized with respect to $\Delta_x$, $\Delta_y$,
and $\theta$; the minimization with respect to $\theta$ gives $\theta = \pm \pi/2$. This minimization gives a heat capacity jump at $T_{\text{c},0}$ of
\[
\frac{\Delta C_0}{T_{\text{c},0}} = \frac{a_0^2}{u + \beta - \alpha}.
\]
When $\sigma \neq 0$, the transition splits into two transitions,
\begin{eqnarray*}
T_\text{c} & = & T_{\text{c},0} + \frac{g|\sigma|}{a_0}, \\
T_\text{TRSB} & = & T_{\text{c},0} - \frac{g|\sigma|}{a_0} \frac{u + \beta - \alpha}{u - \beta + \alpha}.
\end{eqnarray*}
The heat capacity jumps at these two transitions are
\begin{eqnarray*}
\frac{\Delta C_1}{T_\text{c}} & = & \frac{a_0^2}{2u}, \\
\frac{\Delta C_2}{T_\text{TRSB}} & = & \frac{a_0^2}{2u} \frac{u - \beta + \alpha}{u + \beta - \alpha},
\end{eqnarray*}
where $\Delta C_1$ is the heat capacity jump at $T_\text{c}$, and $\Delta C_2$ that at $T_\text{TRSB}$. These
expressions give:
\[
\frac{\Delta C_1}{T_\text{c}} \frac{dT_\text{c}}{d|\sigma|}  = -\frac{\Delta C_2}{T_\text{TRSB}} \frac{dT_\text{TRSB}}{d|\sigma|} =  \frac{a_0 g}{2u}.
\]
In other words, in the mean field approximation the ratio of the slopes $dT_\text{c}/d|\sigma|$ and
$dT_\text{TRSB}/d|\sigma|$ is inverse to the heat capacity jumps $\Delta C/T$ at the transitions.

\subsection{More information on the sample and cell configuration}

We provide here a brief description of the apparatus, and more information on the sample configuration. The
uniaxial stress apparatus is explained fully in Ref.~\cite{Cell18}. To facilitate sample exchange, samples are
mounted in detachable holders that slot into a piezoelectric-driven device that we term the generator. The
generator contains a compression spring that generates a preload force of $\sim$1000~N. This preload is
applied to two sets of piezoelectric actuators. One set is anchored to the cell frame and the other to the
sample. By adjusting the relative lengths of the actuators through the applied voltages, the fraction of the
preload that is applied to the sample is varied. The cell is designed such that it can only apply compressive
forces; it incorporates a mechanical interface that opens when the applied displacement becomes tensile. The
generator incorporates a strain gauge bridge that is configured to measure the displacement applied to the
sample holder. This sensor serves as a complement to the strain gauge-based force sensor incorporated into the
holder. (The strain gauges were supplied by Tokyo Sokki, part number CFLA-3-350-11.) %part number X.  If the
force-displacement relation is linear then the deformation of the sample, and epoxy anchoring the sample, is
elastic. On the other hand a sharp increase in displacement accompanied by a decrease in force indicates that
a fracture or slip occured somewhere.

\begin{figure}
\includegraphics[width=140 mm]{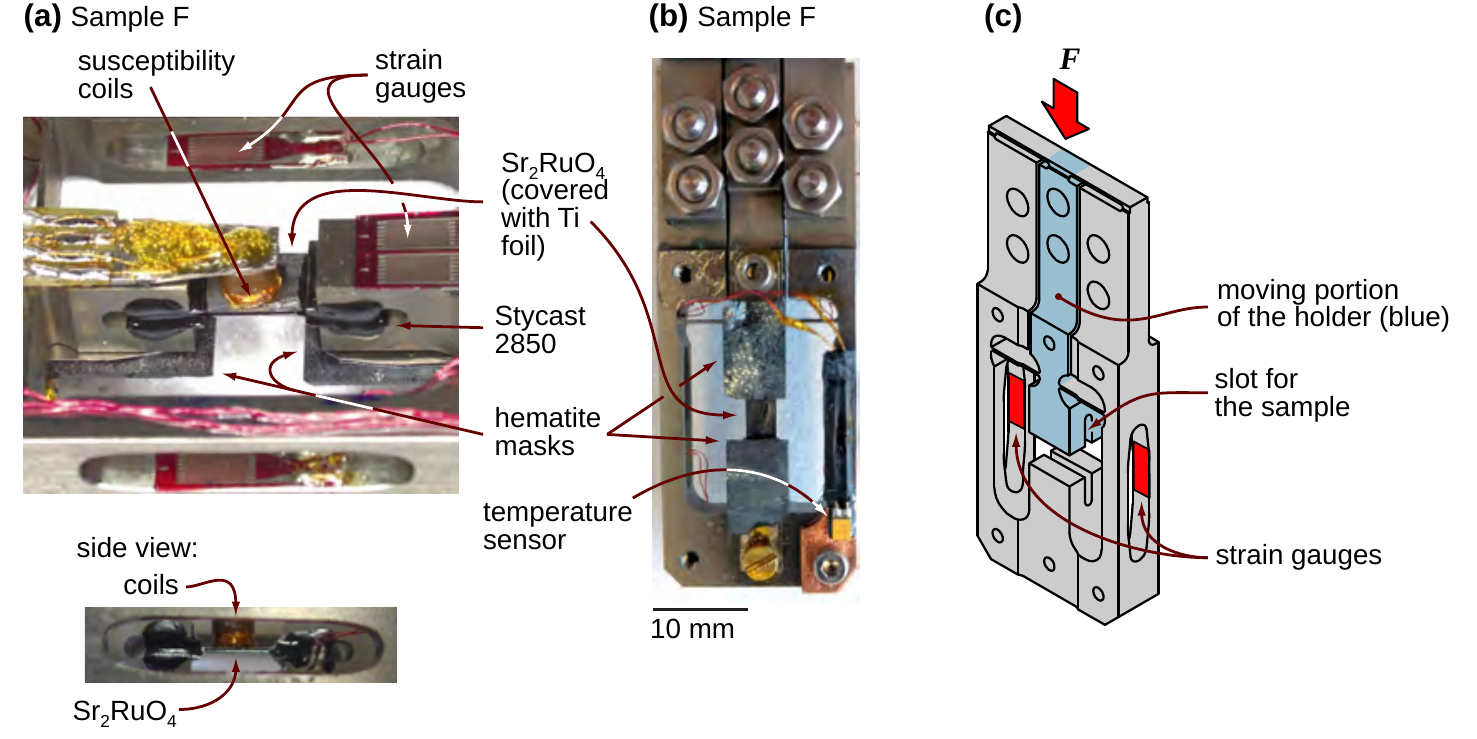}
\caption{\textbf{(a)} Photographs of sample F, showing the location of the ac susceptibility coil, the
hematite masks, and the force-sensing strain gauges. \textbf{(b)} A view from the muon direction for the same
sample. The beam diameter is $\sim$1~cm. \textbf{(c)} Schematic of the sample holder showing the moving portion of the holder (blue color),
slots for the sample and the strain gauges to measure the applied force. The other strain gauges visible in
panel (a) are used to complete the Wheatstone bridge.}  
\label{FigS1}
\end{figure}

Fig.~\ref{FigS1} shows a closer view of sample F, as mounted for measurement. As has been described in the
main text and above, there is a pair of coils for in situ measurement of $T_\text{c}$ through magnetic
susceptibility. The two coils each have $\approx$100~turns, and are wound directly on top of each other onto
the same form. Together, they had an inner diameter of $\approx$1.8~mm and an outer diameter of
$\approx$2.8~mm, and extended from $\approx$0.4 to 1.0~mm above the sample. The measurement is of the mutual
inductance of the two coils using a lock-in amplifier, and the excitation field applied at the sample surface
was typically $\sim$10~$\mu$T.

\begin{figure}[H]
	\includegraphics[width=160mm]{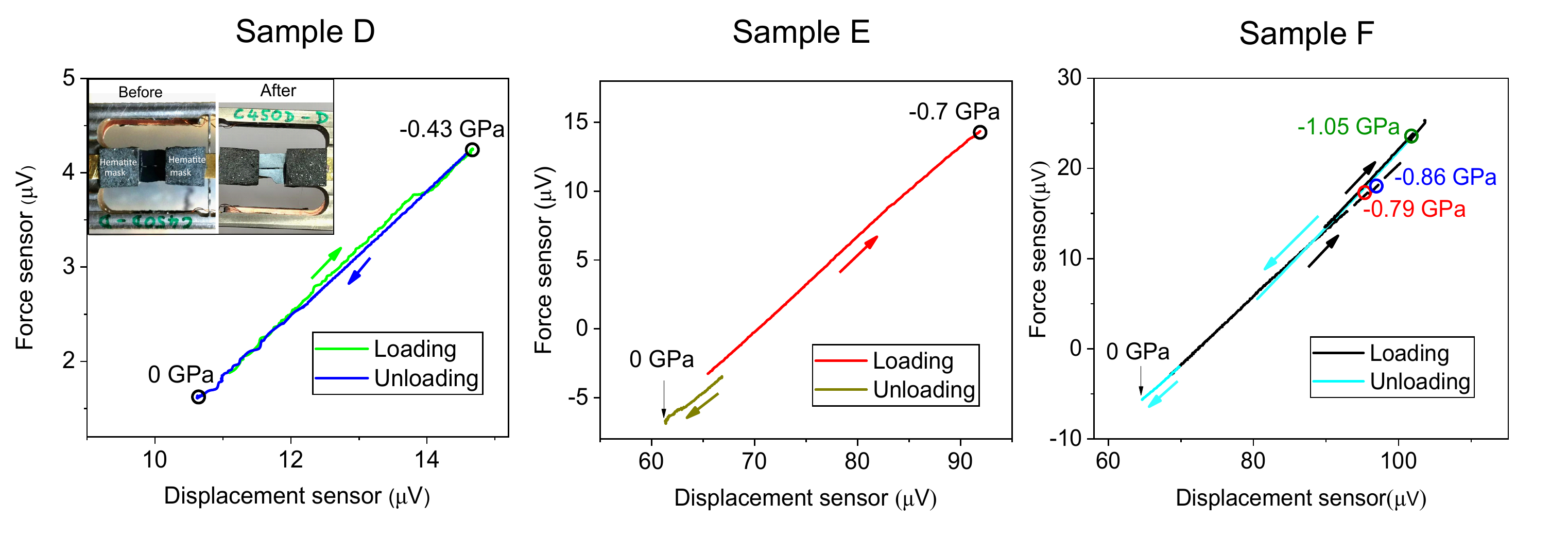}
	\caption{\textbf{Loading curves}: In situ force and displacement sensor readings. A linear
dependence indicates elastic sample deformation. An anomaly in the loading curve for Sample D is where a chip
broke from the sample. Circles indicate the positions of $\mu$SR measurements; stress values were obtained
from the known stress dependence of $T_\text{c}$, and negative values indicate compression.}
	\label{Fig_loading}
\end{figure}

The loading curves, meaning the readings from the force and displacement sensors, for samples D--F are shown
in Fig.~\ref{Fig_loading}. All samples were cooled with the piezoelectric actuators grounded. The readings of
the sensors with no force on the sample could be straightforwardly determined by shifting the actuators
towards tensile force: when the force reaches zero and the mechanical interface opens, the sensor readings
plateau (0 GPa label in Fig.\ \ref{Fig_loading}). All samples were under modest compression after initial
cool-down; we speculate that this is due to differential thermal contraction between loaded and unloaded
piezoelectric actuators. For sample D, there is a prominent anomaly in the loading curve between -0.28 and
-0.43~GPa, which is where a chip broke from the sample (see inset of Fig.~\ref{Fig_loading}). The
chip will have reduced the homogeneity of the applied stress, however the susceptibility data in Fig.~3(h)
show that in spite of it the superconducting transition at -0.43~GPa was not excessively broad.  The
loading curves for samples E and F, in contrast, show no such anomalies. The loading data for Sample F fall on
two distinct lines. We speculate that this was caused by a slip at the holder-generator interface, or
elsewhere in the generator, that shifted the calibration of the displacement sensor. After pulling the sample
out of the system no visible signs of damage to the sample were found.

The force sensor was calibrated for each sample using the known stress dependence of
$T_\text{c}$~\cite{Barber19VHS}. For Samples D and F, $T_\text{c}$ was identified as the midpoint of the
transition in susceptibility; due to the width of the superconducting transition near the Lifshitz transition,
only data with $|\sigma| < 0.53$~GPa were used. Sample E was tuned to the maximum $T_\text{c}$, reported in
Ref.~\cite{Barber19VHS} to occur at -0.70~GPa.

\begin{figure}[H]
	\includegraphics[width=180mm]{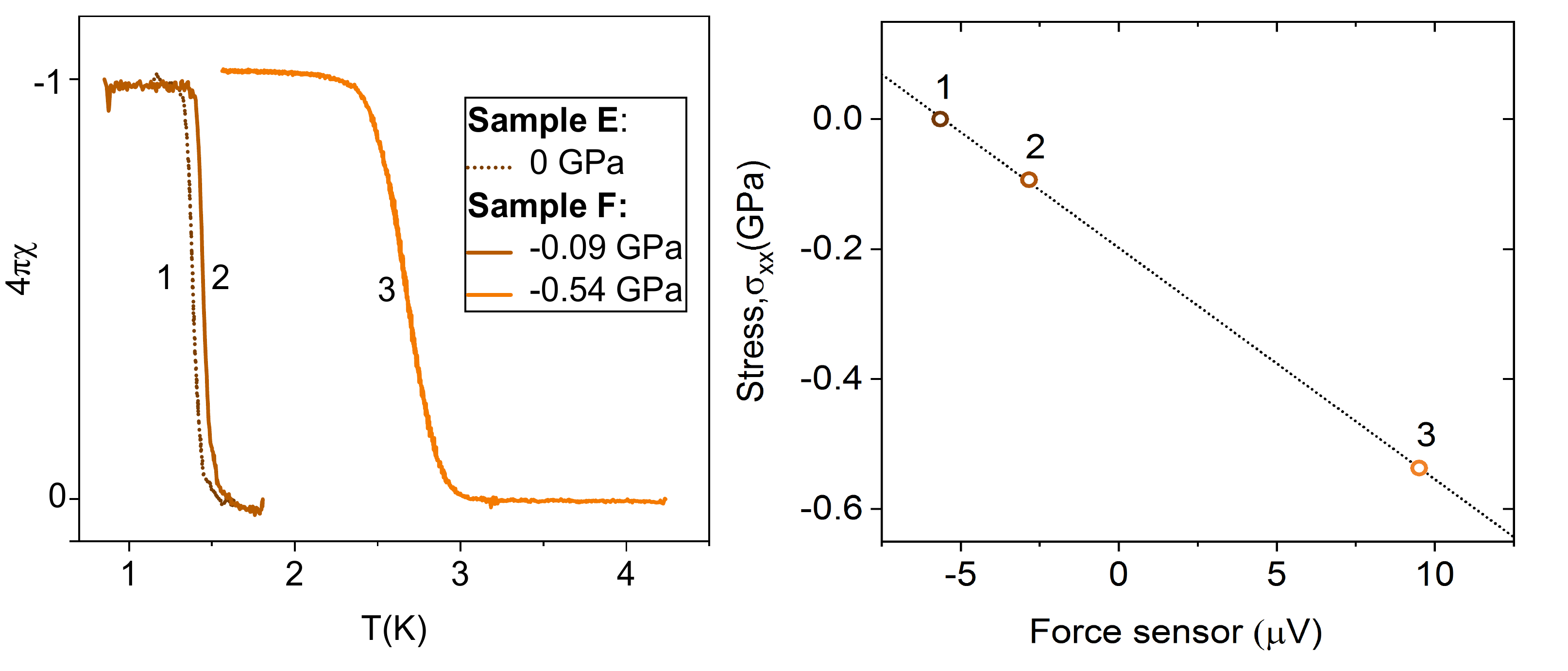}
	\caption{\textbf{Calculating the stress}:  
	(Left panel) ACS curves measured in-situ at different stress values below the peak in $T_c$ for Sample
E and F. (Right panel) Force sensor calibration for sample F. The 0 GPa value is taken from $T_\text{c}$
of Sample E, which was immediately adjacent to Sample F in the original rod.}
	\label{Fig_calib}
\end{figure}

\subsection{Sample characterization}

\begin{figure}[ptb]
\includegraphics[width = 164mm]{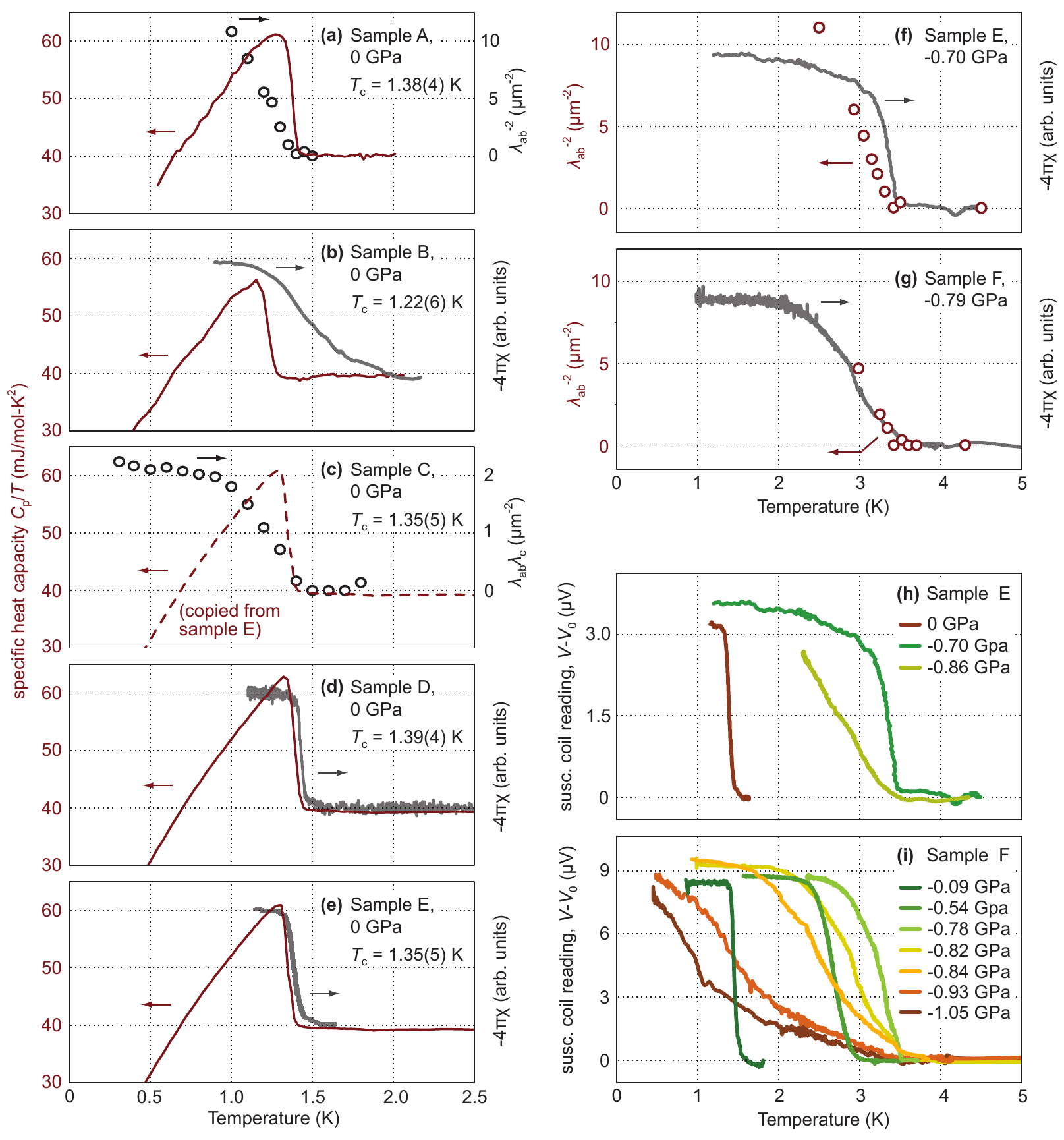}
\caption{\label{fig_heatCap} \textbf{Further characterization of the samples:} All susceptibility data were recorded in-situ. Penetration depths were determined through transverse-field $\mu$SR. Heat capacity data for Samples C, D
, and E were recorded from portions of the samples that were removed from the holder after the $\mu$SR measurement, for Sample A the measurement was performed on a portion of the sample before the $\mu$SR measurement. Note that for Sample C,
we show heat capacity data from Sample E, which was extracted from Sample C. Panels {\bf (h)} and {\bf (i)} show ac susceptibility
 raw voltage data at different uniaxial stresses without normalization.}
\end{figure}

Additional characterization data of the samples are shown in Fig.~\ref{fig_heatCap}. Heat capacity data are
shown for all samples except Sample F; Samples E and F were both drawn from Sample C, and so are expected to
have very similar properties. The samples have sharp transitions with $T_\text{c}$ near the clean-sample limit
of 1.50~K in all samples except Sample B. For Samples D and E, $T_\text{c}$ and the transition widths obtained
from heat capacity and susceptibility data match closely, indicating high sample homogeneity, while for Sample
B the transition in susceptibility is broad and at a higher temperature than the heat capacity transition,
most likely due to internal defects such as Ru inclusions, which locally increase $T_\text{c}$ through strain
effects.

To obtain a superconducting penetration depth we performed transverse field measurements at different
temperatures similar to shown in Fig. \ref{TFdata2}. In the superconducting state the data were analyzed by a
multi-Gaussian fit to obtain the second moment of the internal field distribution in the vortex state as
described in Ref.\ \cite{Maisuradze2009}. The penetration depth $\lambda$ was calculated using Brandt relation
between the magnetic penetration depth and the second moment $< \Delta B^{2}> $ = $ 0.00371 \, \Phi_0^2 /
\lambda^4$, where $\Phi_0$ in the magnetic flux quantum, valid for the case when the applied field $B <<
B_{\rm c2}$  \cite{Brandt2003}. The latter holds for the data presented in Fig.\ref{fig_heatCap}, where the
inverse squared penetration depth $\lambda^{-2}$ is plotted. The measurements shown in panel (a), (f), and (g)
were performed in B$\parallel{c}$ = 2 mT and in panel (c) in B$\parallel{ab}$ = 14.5 mT. Like the heat
capacity data of Ref.~\cite{Li19}, these measurements show definitively that the strain-induced increase in
$T_\text{c}$ is a bulk effect.

Finally, we summarize all measured in-situ susceptibility data for samples E and F in Fig.~\ref{fig_heatCap} {\bf (h)} and {\bf (i)} .
Note that in these panels the raw Lock-In-amplifier voltage data are shown without normalization, the offset voltage $V_0$ above $T_\text{c}$ being subtracted from each curve.  These
data demonstrate that at all measured uniaxial stress values the superconducting volume fraction at lowest temperatures is $\approx$~1.
%The Lifshitz transition is indicated by a sharpening %of the superconducting transition as seen in %susceptibility. 

\bibliography{draft_200206}

\end{document}